# Strong electron-phonon coupling and multiband effects in the superconducting *β*-phase Mo$_{1-x}$Re$_x$ alloys

Shyam Sundar[1], L S Sharath Chandra[2], M K Chattopadhyay[1,2], Sudhir K Pandey[3], D Venkateshwarlu[4], R Rawat[4], V Ganesan[4] and S B Roy[1,2]

[1] Homi Bhabha National Institute at RRCAT, Indore, Madhya Pradesh-452013, India
[2] Magnetic and Superconducting Materials Section, Raja Ramanna Center for Advanced Technology, Indore, Madhya Pradesh-452013, India
[3] School of Engineering, Indian Institute of Technology Mandi, Kamand, Himachal Pradesh-175005, India
[4] UGC-DAEConsortium for Scientific Research, Khandwa Road, Indore, Madhya Pradesh-452001, India

**Abstract**

Superconducting transition temperature T$_C$ of some of the cubic β-phase Mo$_{1-x}$Re$_x$ alloys with x > 0.10 is an order of magnitude higher than that in the elements Mo and Re. We investigate this rather enigmatic issue of the enhanced superconductivity with the help of experimental studies of the temperature dependent electrical resistivity (ρ(T)) and heat capacity (C$_P$(T)), as well as the theoretical estimation of electronic density of states (DOS) using band structure calculations. The ρ(T) in the normal state of the Mo$_{1-x}$Re$_x$ alloys with $x \geq 0.15$ is distinctly different from that of Mo and the alloys with $x < 0.10$. We have also observed that the Sommerfeld coefficient of electronic heat capacity γ, superconducting transition temperature T$_C$ and the DOS at the Fermi level show an abrupt change above $x > 0.10$. The analysis of these results indicates that the value of electron–phonon coupling constant λ$_{ep}$ required to explain the T$_C$ of the alloys with $x > 0.10$ is much higher than that estimated from γ. On the other hand the analysis of the results of the ρ(T) reveals the presence of phonon assisted inter-band s–d scattering in this composition range. We argue that a strong electron–phonon coupling arising due to the multiband effects is responsible for the enhanced T$_C$ in the β- phase Mo$_{1-x}$Re$_x$ alloys with $x > 0.10$.

**Introduction**

Superconducting transition temperature *T$_C$* of several transition metal binary alloys such as Ti$_x$V$_{1-x}$, Nb$_{1-x}$Zr$_x$, and Mo$_{1-x}$Re$_x$, etc., is higher than that of the constituent elements themselves [1]. Among them, Mo$_{1-x}$Re$_x$ and Zr$_{1-x}$Rh$_x$ alloys have a distinction that the *T$_C$* is about an order of magnitude higher in these alloys in comparison with the constituent elements [1–3]. In the case of Zr$_{1-x}$Rh$_x$ alloys, such an enhancement of *T$_C$* happens in a new crystallographic structure as observed in several Nb and V based compounds such as Nb$_3$Sn, Nb$_3$Ge, V$_3$Si, and V$_3$Ge etc. However, the class of β-phase Mo$_{1-x}$Re$_x$ alloys is quite distinct, where the order of magnitude increase in *T$_C$* takes place within the same structure of one of the parent elements (Mo in this case). These Mo–Re alloys possess excellent mechanical properties at elevated temperatures and find widespread applications in aerospace and defence industries, medical fields and welding production [4–7]. The variation of *T$_C$* in Mo$_{1-x}$Re$_x$ alloys with the increase in the concentration *x* is non monotonic in nature; the *T$_C$* increases slowly from 0.90 K for Mo to about 3 K for *x* = 0.10 and then rises sharply to about 7 K for *x* = 0.15 [8]. With further increase in *x*, the *T$_C$* increases linearly to about 12.6 K for *x* = 0.40 [8]. The composition range in which the

$T_C$ increases sharply in the Mo$_{1-x}$Re$_x$ alloys, corresponds to the same composition range where the existence of two electronic topological transitions (ETT) with the corresponding critical concentrations $x_{C1}$ = 0.05 and $x_{C2}$ = 0.11 have been reported in the literature [9–15]. The ETT is associated with the appearance or the disappearance of pockets of Fermi surface when an external parameter such as composition, pressure, and/or magnetic field is varied [16]. The experimental evidences for the ETT in the Mo$_{1-x}$Re$_x$ alloys were obtained in the form of a giant enhancement in the thermoelectric power [9, 12, 13], oscillations in the $T_C$ and critical magnetic field $H_C$ as a function of pressure for $x_{C2}$ = 0.11 [13]. More direct evidence of ETT in this system has been provided recently by the angle resolved photo-emission spectroscopy along the H–N direction of the Brillouin zone [15]. Theoretically, the elastic constants are also expected to show non monotonic variation across $x_{C1}$ and $x_{C2}$ [10], though no such signature is observed experimentally in the Mo$_{1-x}$Re$_x$ alloys so far [17]. However, softening of the lattice resulting in the appearance of the Brout–Visscher local phonon resonance mode [18] of Re is observed in tunnelling spectra [19, 20] of the alloys with $x > x_{C2}$.

Shum et al [19] provides an explanation for the enhancement of TC in the Mo 0.60 Re 0.40 alloy by considering a Brout–Visscher local phonon mode that contributes strongly to the electron–phonon coupling function α2F (ω). However, the results of subsequent point contact spectroscopy study of quasi-local vibrations in the Mo1−xRex alloys pointed out that the lattice softening alone could not possibly explain the enhancement of TC in these alloys [20]. The lattice softening is actually influenced by two factors [20]. First, the doping with relatively heavy element Re, which leads to the appearance of quasi-local vibrations or Brout–Visscher mode [19]. Secondly, electron density of states (DOS) at the Fermi level increases considerably with Re doping, which in turn decreases the force constants and renormalizes the electron–phonon interaction. Thus the electronic factor N(0)<I$^2$⟩ (where N(0) is the electron DOS at the Fermi level EF and <I$^2$>is the matrix element of the electron–phonon coupling) also plays a significant role in enhancing the T$_C$ [20].

A clear cut relationship between the ETT and the superconducting properties, however, is yet to be established in the Mo$_{1-x}$Re$_x$ alloys [15]. In a recent study on the temperature dependence of lower critical field and heat capacity, we have shown the existence of the multiband effects in the superconducting state of the Mo$_{1-x}$Re$_x$ alloys with $x$ = 0.25 and 0.40 [21]. We have suggested that the Fermi pockets of Re 5d like character appear in those alloys and contribute to the superconductivity in a distinct manner than the rest of the bands [21].In this paper, we present a study of the normal state and superconducting properties of the Mo$_{1-x}$Re$_x$ alloys with $0 \leqslant x \leqslant 0.4$, performed through the measurements of the temperature dependence of the resistivity and heat capacity in different magnetic fields. We show that for the alloys with $x < 0.1$, the intra-band s–s interaction is predominant, whereas for the alloys with $x > 0.1$ phonon assisted inter-band s–d interaction becomes a major source of scattering. Our studies also reveal that the phonons involved in the inter-band s–d scattering are of low energy, and arise due to the softening of the lattice. With the help of these experimental results and a theoretical estimation of electronic DOS using band structure calculations, we then argue that the enhancement of $T_C$ in the Mo$_{1-x}$Re$_x$ alloys for $x \geqslant 0.15$ can be explained within a framework of the enhanced electron-phonon coupling through the phonon assisted inter-band s–d interaction.

**Experimental**

Polycrystalline samples of $Mo_{1-x}Re_x$, where (x = 0, 0.025, 0.05, 0.075, 0.1, 0.15, 0.20, 0.25, and 0.40) were prepared by melting 99.95+% purity constituent elements in an arc furnace under 99.999% Ar atmosphere. The samples were flipped and re-melted six times to improve the homogeneity. The x-ray diffraction study of these alloys shows that the samples have formed in the body centred cubic (bcc) phase (space group: Im 3m). The heat capacity measurements were performed in the temperature range 2–200 K in various magnetic fields up to 3 T using the Physical Property Measurement System (PPMS, Quantum Design, USA). The resistivity measurements in zero field were performed using the standard four probe configuration in the temperature range 2–300 K using 9 T cryostats (Oxford Instruments, UK and PPMS, Quantum Design, USA).

The ab initio electronic structure calculations were performed using the spin polarized Korringa–Kohn– Rostoker (KKR) method [22]. The effect of doping was considered under the coherent potential approximation. The exchange correlation functional developed by Vosko, Wilk and Nusair was used for the calculation [23]. The number of k-points used in the irreducible part of the Brillouin zone is 72. The muffin-tin radii for Mo and Re atoms used in the calculations are same and equal to 2.576 bohr. For the angular momentum expansion, we have considered $l_{max}$ = 2 for each atom. The potential convergence criterion was set to $10^{-6}$.

**Results and discussion**

*1. Resistivity of $Mo_{1-x}Re_x$ alloys in their normal state*

Figures 1 shows the temperature dependence of resistivity ρ(T) for the $Mo_{1-x}Re_x$ alloys. The open symbols are the experimental data points. The ρ(T) for elemental Mo is in agreement with the literature [1, 24] and is similar to several other elemental metals such as Cu, Au, and Ag etc., [25]. When Re is alloyed with the Mo, the nature of the temperature dependence of resistivity ρ(T) gradually changes, and for x = 0.15–0.40, the form of ρ(T) is similar to that typically observed in the A-15 compounds such as $Nb_3Sn$, and $V_3Si$ [26, 27]. The temperature dependence of resistivity in such cases can be fitted to empirical relation [26, 27]

$$\rho(T) = \rho_0 + \rho_1 T^n + \rho_2 \exp(-T_0/T),  \qquad [1]$$

where $\rho_0$ is the residual resistivity, $\rho_1$, $\rho_2$, $n$ and $T_0$ are constants. The temperature $T_0$ is related to certain thermally activated phonon process [28, 29]. The solid lines in figure 1 show the fit obtained using above equation for $x \geqslant 0.15$.

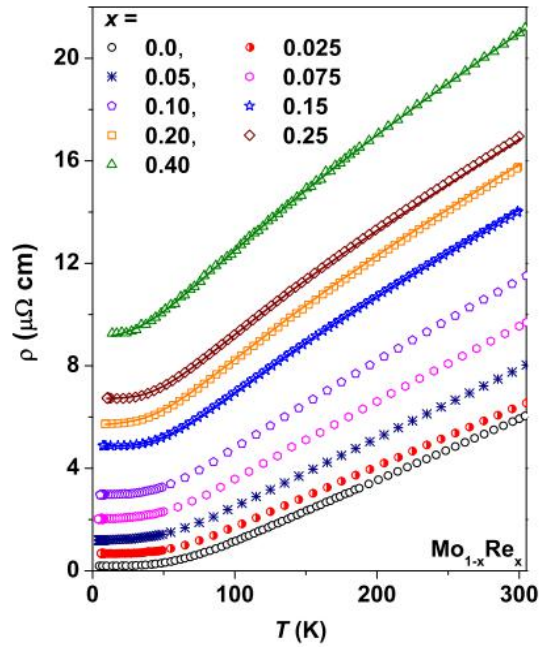

Figure 1. Temperature variation of resistivity ρ(T) of the Mo$_{1-x}$Re$_x$ alloys for 0 $x$ 0.40. The open symbols are the data points and the solid lines are the fit to the data for $x$ 0.15 using equation (1). The temperature dependence of resistivity for the alloys with x > $x_{C2}$ = 0.11 is distinctly different from that of molybdenum (x = 0) and alloys with low Re composition.

## 2. Physical Properties of Mo$_{1-x}$Re$_x$ alloys below the superconducting transition temperature $T_C$

Figure 2(a) shows the ρ(T) around the superconducting transition temperature $T_C$. The residual resistivity $ρ_0$ (which is the resistivity just above the $T_C$) increases with increasing x. The values of $ρ_0$ for x = 0.15 and 0.25 are in agreement with those reported earlier for the same compositions [30]. The $T_C$ is estimated from ρ(T) as the temperature at which temperature derivative of the ρ(T) shows the maximum. The temperature dependence of heat capacity C(T) in zero magnetic field is shown in figure 2(b) for the alloys that are superconducting above 2 K (0.1 $x$ 0.4). The $T_C$ is estimated from C(T) as the temperature at which the temperature derivative of the C(T) shows the minimum.

Figure 3(a) shows the temperature dependence of C in the normal state of the alloys plotted as C/T versus T$^2$ in the temperature range 2–10 K. The normal state is prepared by the application of magnetic fields sufficient (up to 30 kOe) to suppress the superconducting transition in these alloys to <2 K. The solid lines are straight line fit to the experimental data (symbols) obtained using the relation C/T = γ + βT$^2$. Here, γ=(π$^2$/3) k$^2_B$ N(0) is the Sommerfeld coefficient of electronic specific heat where k$_B$ is the Boltzmann constant. The constant β is related to Debye temperature $θ_D$ as $θ_D$=(1943.66/β)$^{1/3}$. The value of C/ γ$T_C$ is obtained by estimating the jump C in the temperature dependence of heat capacity at $T_C$. The value of C/ γ$T_C$ is about 1.46 for x = 0.1 and increases further to reach to a value of 2.08 for x = 0.40. Figure 3(b) shows the variation of $T_C$ estimated from ρ(T) as well as C(T) along with the γ as a function of Re content x. The errors in the estimation of $T_C$ from resistivity is less than ±0.07 K and that from heat capacity is less than ±0.2 K, while the error in estimation of γ is less than 1%. The value of $T_C$ for Mo is taken from the [1]. The $T_C$ estimated from both the measurements are in agreement within 0.2 K. The $T_C$ appears to be increasing linearly (the dashed green line which is a guide to eye) from 0.9 K for x = 0 to about 3 K for x = 0.1. The $T_C$ rises sharply above x = 0.1 and is linear above x = 0.15. It is also interesting to note from figure 3(b) that the compositional dependence of γ in $Mo_{1-x}Re_x$ alloys follows the compositional dependence of $T_C$. The sharp rise in $T_C$ as well as in γ takes place in the same Re concentration region where an ETT in the $Mo_{1-x}Re_x$ alloys has earlier been reported [9–15]. The Debye temperature $θ_D$ on the other hand decreases linearly with increasing x. The enhancement of γ generally signifies the enhancement in the DOS at the Fermi level. At the first glance, it can be inferred that the $T_C$ in the $Mo_{1-x}Re_x$ alloys is governed mainly by the DOS at the Fermi level. However, this statement should be taken with some caution as the electron–phonon coupling also plays an important role in determining the $T_C$ [31]. This is supported by the fact that the value of $C/γT_C$ for $Mo_{1-x}Re_x$ alloys is higher than the Bardeen Cooper and Schrieffer (BCS) theoretical limit of 1.43 [32].

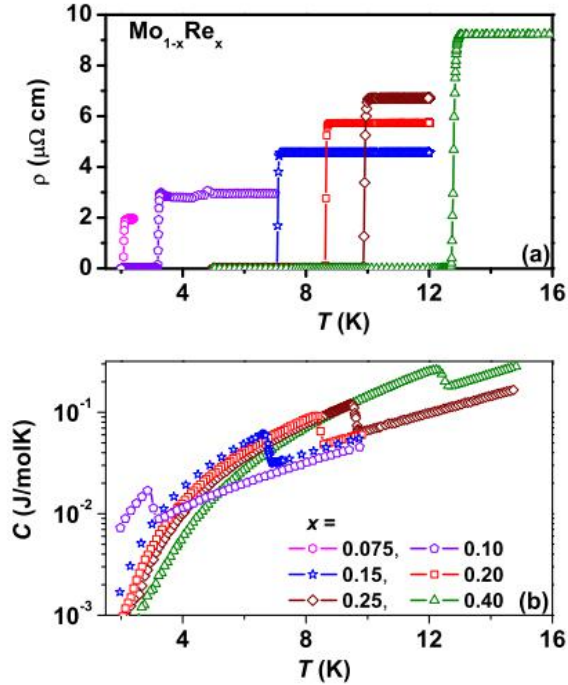

Figure 2. (a) Temperature variation of ρ around $T_C$ of the $Mo_{1-x}Re_x$ alloys that are superconducting above 2 K. The residual resistivity $ρ_0$, which is the resistivity just above $T_C$, increases with increasing x. (b) The temperature dependence of heat capacity C(T) in zero magnetic field of the alloys that are superconducting above 2 K.

## 3. Electronic structure of $Mo_{1-x}Re_x$ alloys

In order to obtain a proper understanding of the roles of electron DOS and electron-phonon coupling in the superconductivity of the $Mo_{1-x}Re_x$ alloys, we have performed the electronic structure calculations and obtained the electronic DOS as a function of energy (figure 4(a)). The Fermi level of Mo and the low x alloys lies around a valley in the DOS and there is a small peak just above the Fermi level. This resembles the DOS of the pseudo gap systems such as CoSi etc [33]. The energy dependence of the DOS for the alloy compositions is similar to that of Mo. However, at larger Re concentrations, the fine features are smeared out due to disorder effects.

When Re is alloyed with Mo, the energy dependence of the DOS shifts to negative (lower) energies like that in a rigid band model due to charge transfer to Mo from Re. The variation of DOS at the Fermi level $E_F$ as a function of x is plotted in figure 4(b). We have also drawn two marker lines corresponding to $x_{C1} = 0.05$ and $x_{C2} = 0.11$ which highlight the changes in the DOS at the Fermi level around the ETT. The DOS at $E_F$ decreases slightly for x = 0.025 as compared to x = 0 and then increases sharply as x is increased up to x = 0.10. For $x$ 0.10, the DOS at $E_F$ increases with x, but with a much slower rate as compared to $x$ 0.10. The DOS at the Fermi level shows a drop again for x = 0.40. The compositional dependence of DOS at $E_F$ does not follow the compositional dependence of $T_C$ and $\gamma$ (figure 3(b)). Therefore, it becomes important to investigate the possible influence of the electron–phonon coupling constant $\lambda_{ep}$ on the $T_C$ of the system.

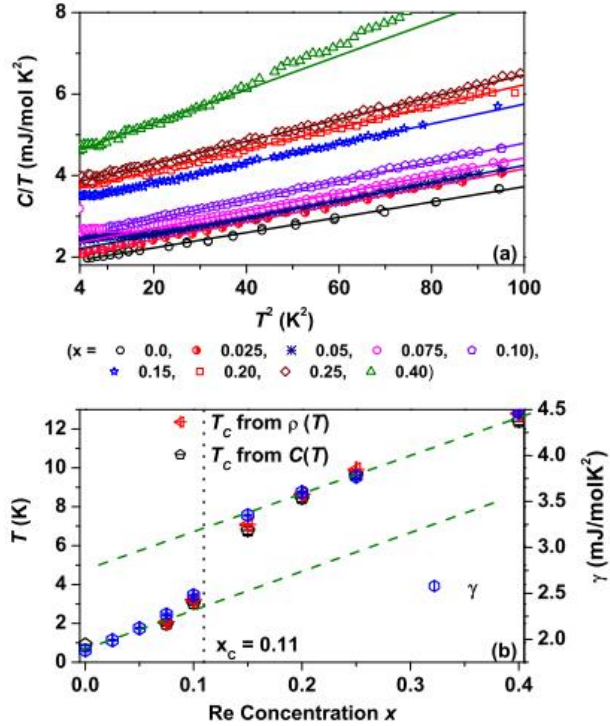

Figure 3. (a) Temperature dependence of heat capacity C(T) in the normal state plotted as C/T versus $T^2$ in the temperature range 2–10 K. The open symbols are the experimental data points and the fitted straight lines represent the relation $C/T = \gamma + \beta T^2$. (b) The variation of $T_C$ estimated from both $\rho(T)$ as well as C(T) as a function of x. The $T_C$ estimated from both the measurements are in agreement within 0.2 K. The variation of $\gamma$ with respect to x is also compared with the $T_C(x)$. The dotted lines are guide to eye. The value of x at which both $T_C$ and $\gamma$ increases sharply also corresponds to the critical concentration for the electronic topological transition.

The electron–phonon coupling constant can be estimated from the $\gamma$ as [34, 35]

$$\lambda_{ep}^\gamma = \frac{\gamma}{\gamma_0} - 1,$$

[2]

where $\gamma_0$ is the Sommerfeld coefficient of specific heat estimated from the DOS at $E_F$ obtained from the band structure calculations. In the case of strong coupling superconductors, the $\lambda_{ep}$ can also be estimated from $C/\gamma T_C$ and $T_C$ as [31]

$$\frac{\Delta C}{\gamma T_C} = 1.43\left[1 + 53\left(\frac{T_C}{\omega_{\ln}}\right)^2 \ln\frac{\omega_{\ln}}{3T_C}\right]$$

[3]

where $\omega_{\ln}$ is the average phonon frequency, and

$$T_C = \frac{\omega_{\ln}}{1.2} \exp\left(\frac{1.04(1 + \lambda_{ep}^{sc})}{\mu^* + 0.62\mu^*\lambda_{ep}^{sc} - \lambda_{ep}^{sc}}\right).$$

[4]

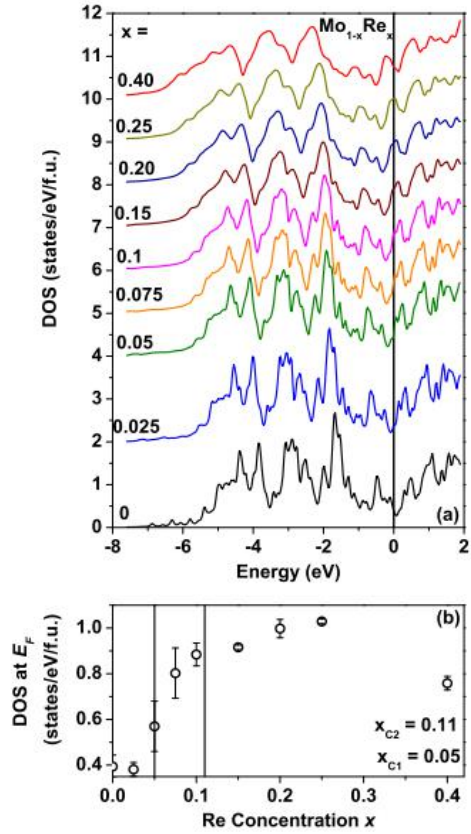

Figure 4. (a) Electronic density of states of the $Mo_{1-x}Re_x$ alloys estimated using the KKR method. The Fermi level of Mo lies just below a valley in the density of states which has a small peak just above the Fermi level. When Re is alloyed with Mo, the Fermi level ($E = 0$) shifts to higher energies like that in a rigid band model due to charge transfer to Mo from Re. (b) The variation of DOS at $E_F$ as a function of $x$. We have also drawn two marker lines corresponding to $x_{C1} = 0.05$ and $x_{C2} = 0.11$ which highlights the changes in the DOS at $E_F$ around the ETT.

The $\lambda_{ep}$s ($\lambda_e^{\gamma}{}_p$, $\lambda^{sc}{}_{ep}$) estimated with these methods are plotted as a function of Re concentration x in figure 5. In the case of Mo, the $\lambda_{ep}$ is estimated from the McMillan formula $T_C = (\theta_D/1.45)\exp(1.04(1+\lambda_{ep})/(\mu^*+0.62\ \mu^*\lambda_{ep}\ -\lambda_{ep}))$ [36] instead of the strong coupling formula. At very low compositions (x < 0.075), we could not estimate the $\lambda^{sc}{}_{ep}$ using C/ $\gamma T_C$ and $T_C$ as their $T_C$ is less than 2 K, which is beyond the range of our present experimental set-up. The values of $\lambda_e^{\gamma}{}_p$ for all the alloys do not match with $\lambda^{sc}{}_{ep}$. In the case of Mo, the value of $\lambda^{sc}{}_{ep}$ is smaller than $_e{}^{\gamma}{}_p$. We had observed similar behaviour in the Ti–V binary alloys as well, and pointed out that the spin fluctuations were responsible for such effect [35, 37]. It is to be noted here that any possible influence of spin-fluctuations was not taken into account while estimating $\lambda_{ep}^{\gamma}$ in the present $Mo_{1-x}Re_x$ alloys. In the presence of spin-fluctuations, $\lambda_{ep}^{\gamma}$ is renormalized by a factor $(1 + \lambda_{sf})^{-1}$, where $\lambda_{sf}$ is electron spin-fluctuations coupling constant [35, 37]. On the other hand while estimating $\lambda_{ep}^{sc}$ using $C/\gamma T_C$, the effect of spin-fluctuations is already taken in to account in the experimentally determined value of $T_C$. The theoretical and experimental studies by Hasegawa and Wierzbowska show that the $\lambda_{ep}$ is about 0.4 and the $\lambda_{sf}$ is about 0.04 in molybdenum [38, 39]. Thus the spin fluctuations might have some influence in Mo and $Mo_{1-x}Re_x$ alloys with low Re concentrations. However, we will not probe these alloys any further at present as the superconductivity in these alloys are observed only below 2 K. A detailed experimental study of magnetization, resistivity and specific heat investigating the possible role of spin-fluctuations (if any) in the $Mo_{1-x}Re_x$ alloys will be part of our future study. However, we will argue latter on in the paper that the spin-fluctuations are not playing any dominant role in the $Re_x$ alloys in the composition range x = 0.075–0.25 where the value of $\lambda_{ep}^{sc}$ is larger than $\lambda_{ep}^{\gamma}$. For x = 0.40 on the other hand, the value of $\lambda_{ep}^{sc}$ is smaller than that of $\lambda_{ep}^{\gamma}$.

*4. Soft phonon modes in $Mo_{1-x}Re_x$ alloys and its influence on the superconductivity*

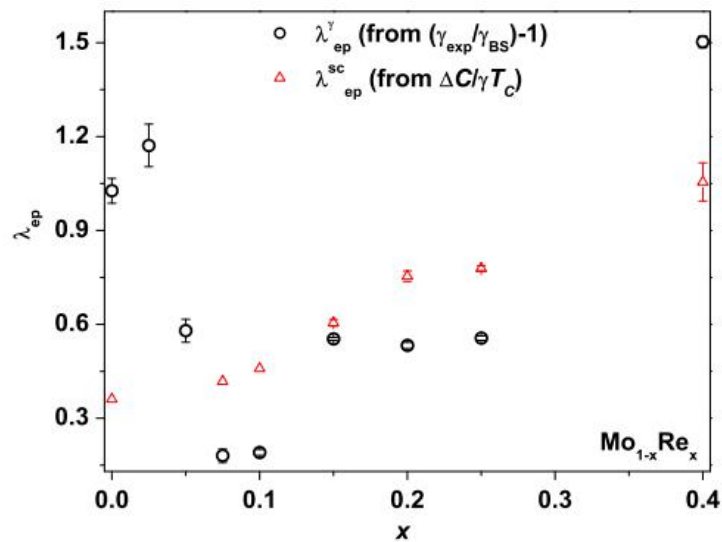

Figure 5. Variation of electron-phonon coupling constant $\lambda_{ep}$ as a function of Re concentration x. The open circles represent the $\lambda_{ep}$ estimated from the experimental $\gamma$ and the DOS at $E_F$ obtained from the band structure calculation. The open triangles represent the $\lambda_{ep}$ estimated from the strong coupling formula for $C/\gamma T_C$ and $T_C$ [31]. In the case of Mo, the $\lambda_{ep}$ is estimated from MacMillan formula instead of strong coupling formula. The $\lambda_{ep}$ estimated from $\gamma$ does not match with that estimated from $C/\gamma T_C$ and/or $T_C$.

The mass defect of Mo and Re i.e., $M_{Re}/M_{Mo}$ is about 1.94, which is quite high. Therefore, the addition of Re in Mo introduces strong disorder in the lattice which disturbs the phonon spectrum and gives rise to the quasi local vibration or the Brout–Visscher mode [18]. This in turn leads to lattice softening [19]. The quasi local mode contributes appreciably to the electron–phonon coupling function $\alpha^2 F(\omega)$ and to $2/k_B T_C$. Shum et al argued that the enhancement of $T_C$ in the $Mo_{0.60}Re_{0.40}$ alloy is due to the lattice softening [19]. The temperature dependence of heat capacity can also provide information on the soft phonon modes [40–42]. The deviation from the linearity of C/T with respect to T² at low temperature is one such signature [43]. It is observed that the range over which this linearity is observed in the C/T versus T² curve, decreases with increasing Re concentration (figure 3(a)). The presence of soft phonon modes can also be inferred by estimating $F(\omega)$ by fitting the temperature dependence of heat capacity using the technique of inverse problem [44]. In this procedure, the temperature dependence of heat capacity is iteratively fitted by a proper choice of trial function and the parameters are refined using least square method [44]. We have used Levenberg–Marquardt algorithm for this fitting procedure [45]. The details of the modeling are given below.

For a system with N atoms per formula unit, there can be 3N modes of vibrations. Out of these modes, three modes are acoustic modes, which can be modelled by Debye theory, whereas 3N-3 optical modes can be modelled by Einstein theory. Therefore, the temperature dependence of lattice heat capacity can be explained by the formula [42, 44]

$$C_L = \int_0^{\omega_{max}} d\omega F(\omega) \frac{e^\omega \omega^2}{(e^\omega - 1)^2}$$

[5]

where $F(\omega)$ is given by

$$F(\omega) = \sum_{i=1}^{3} \frac{24.942 \nu_i \omega^2 Z(\omega - \theta_{Di})}{\theta_{Di}^3} + \sum_{i=1}^{3N-3} A_i \exp\left(\frac{-2^*(\omega - \omega_{E_i})^2}{d_i^2}\right).$$

[6]

Here, $\nu_i$ is the weight factors, $Z(\omega)$ is a step function and $\omega_E$ is the Einstein frequency. The first term corresponds to Debye model and the second term corresponds to Einstein model.

First we have fitted the temperature dependence of heat capacity of Mo in the temperature range 5–50 K. The initial values of various parameters for this fitting are taken from the literature for the $F(\omega)$ of Mo derived theoretically [46]. We have observed that the temperature dependence of heat capacity of Mo in the range 5–50 K can be explained by considering two Debye terms and two Einstein terms. The $F(\omega)$ obtained from the fit closely resembles the $F(\omega)$ available in the literature [46]. This is also in consonance with the phonon spectral function $\alpha^2 F(\omega)$ obtained from tunnelling experiments [20]. We have also observed that the two Debye terms and the two Einstein terms are sufficient to fit the temperature dependence of heat capacity of the $Mo_{1-x}Re_x$ alloys in the range 5–50 K. Figure 6 shows the $F(\omega)$ obtained from the fit using equation (5) to the temperature dependence of heat capacity of $Mo_{1-x}Re_x$ alloys with x = 0–0.40. We have observed that the $F(\omega)$ for alloys up to x = 0.1 is similar to that of Mo. However, phonon spectra shift to lower frequencies, which is an indication of lattice softening. The softening becomes significant for the composition higher than 0.1. The Einstein local mode present just above 20 meV in Mo shifts to less than 15 meV for $x$ 0.15. These phonon spectra are also in line with the $\alpha^2 F(\omega)$ of the $Mo_{1-x}Re_x$ alloys obtained from tunnelling experiments [20].

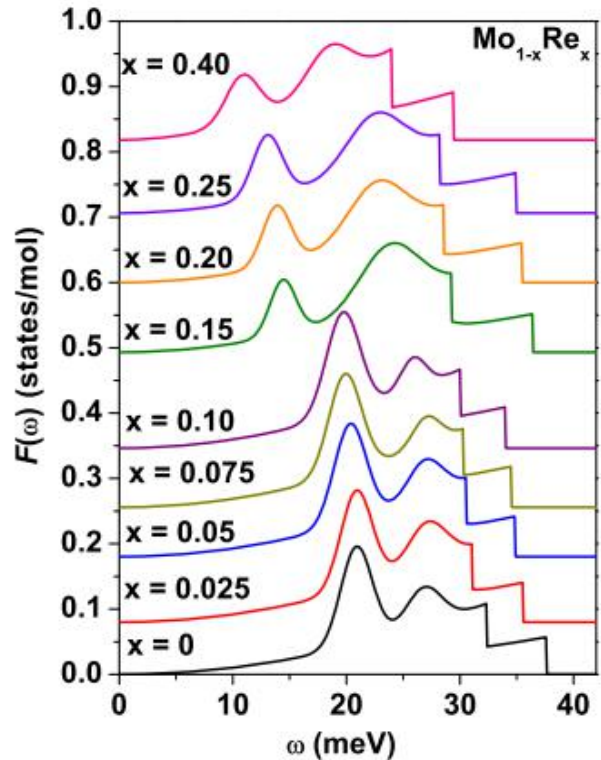

Figure 6. Frequency dependence of phonon spectrum F(ω) for the $Mo_{1-x} Re_x$ alloys obtained from C(T). The F(ω) curves look similar for all the composition up to x = 0.1. The phonon spectrum F(ω) shifts to lower frequencies when x is increased above 0.1 which is an indication of phonon softening.

The average phonon frequency $\omega_{ln}$ can be estimated from F(ω) as [37, 43, 44]

$$\omega_{ln} = \exp\left(\frac{\int F(\omega)\ln\omega\, d\ln\omega}{\int F(\omega)\, d\ln\omega}\right),$$

[7]

Figure 7 shows the average phonon frequency $\omega_{ln}$ expressed in Kelvin as a function of Re composition estimated from the F(ω) obtained from the temperature dependence of normal state heat capacity. The value of $\omega_{ln}$ is about 250 K for Mo and decreases linearly with composition x except for a lateral shift at about $x_{C2}$=0.11. The enhancement of $T_C$ with the decrease of $\omega_{ln}$ with composition indicates the enhancement in the electron–phonon coupling. We have compared the $\omega_{ln}$ obtained from C/$\gamma T_C$ with that estimated from the C(T) in figure 7. The $\omega_{ln}$ estimated from both the methods agree over the range of x = 0.2–0.4. We have also plotted in figure 7, the $T_0$ obtained from the fitting the equation (1) to the ρ(T) for the alloys. It is also observed that the magnitude of $T_0$ is somewhat similar to that of $\omega_{ln}$.

The figure 8 shows the comparison of the experimental $T_C$ with the $T_C$ estimated using $\omega_{\ln}$ obtained from $F(\omega)$, and $\lambda_{ep}^{\gamma}$. The $T_C(x)$ obtained experimentally follows closely the $\lambda_{ep}^{sc}(x)$ curve, whereas the estimated $T_C(x)$ closely follows $\lambda_{ep}^{\gamma}$. For Mo and the $Mo_{1-x}Re_x$ alloys with x up to 0.05, the estimated $T_C$ is much higher than the experimentally obtained one. We had observed such differences in Ti–V alloys as well and we have attributed this to the presence of spin fluctuations in those alloys [35, 37]. In the composition range 0.075–0.25, the experimental $T_C$ is higher than the estimated $T_C$ (shaded region in figure8), and the $\lambda_{ep}$ estimated from $\gamma$ is smaller than that estimated from $C/\gamma T_C$ and $T_C$. Hence spin-fluctuations may not be playing a dominating role in the $Mo_{1-x}Re_x$ alloys in the composition range x = 0.075 − 0.25. This observation is further substantiated by the results of temperature dependent resistivity (presented latter in this paper) showing a very small contribution from the $T^2$ term, which rule out a predominant role of spin-fluctuations in this composition range. Furthermore, the results of our preliminary investigation of normal state dc magnetization do not reveal the typical signature of spin fluctuations as was observed in Ti–V alloys [35, 37]. Tulina and Zaitsev pointed out from the point contact spectroscopy studies that the $\lambda_{ep}$ estimated from the phonon spectral function $\alpha^2 F(\omega)$ alone cannot explain completely the enhancement of $T_C$ in the $Mo_{1-x}Re_x$ alloys [20]. They argued that there must be significant contribution from the electronic factor $N(0)\langle I^2 \rangle$ towards the enhancement of $T_C$ in these alloys [20].

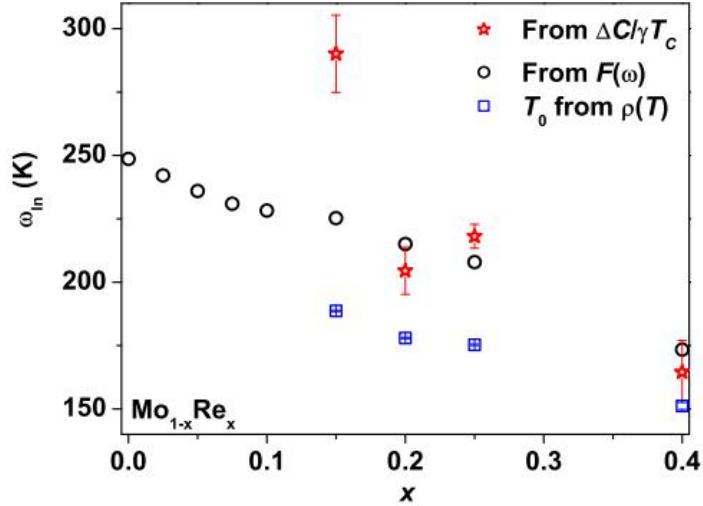

Figure 7. Average phonon frequency $\omega_{\ln}$ expressed in Kelvin as a function of Re content estimated from the strong coupling formula for $C/\gamma T_C$ [31] (open stars) and from the $F(\omega)$ obtained from the temperature dependence of normal state heat capacity.

The lower values of $\lambda_{ep}^{\gamma}$ in comparison with $\lambda_{ep}^{sc}$ are also observed in iron pnictides [47] and $MgB_2$ [48]. Especially in $MgB_2$, this is attributed to the multi-band effect with anisotropic strong inter-band scattering [48]. Our recent studies on the C(T) in the superconducting state of the $Mo_{1-x}Re_x$ alloys revealed that the multiband effects were important in these alloys [21]. Therefore we looked for multiband effects in the normal state properties as well. We have pointed out earlier that the $T_0$ in the equation (1) is related to certain thermally activated phonon process affecting the electrical conduction. In fact, $T_0$ in the case of transition metals and alloys is related to the phonon assisted s–d scattering [28] and it corresponds to the energy of phonons that is assisting the electrons to scatter from s band to d band [29]. This also seems to be true for the present alloys since the magnitude of $T_0$ is similar to that of $\omega_{ln}$ (figure 7) in the present case.

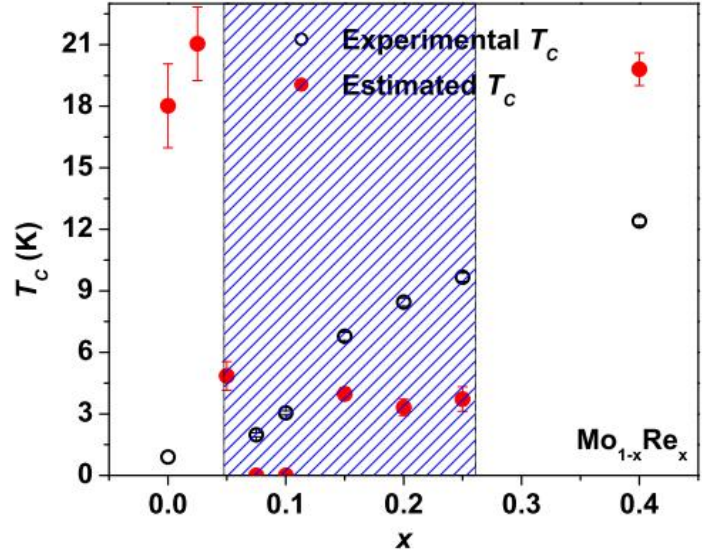

Figure 8. Comparison of $T_C$ obtained experimentally with that estimated using $\omega_{ln}$ obtained from $F(\omega)$ and $\lambda_{e^-p}^\gamma$. For some compositions (shaded region), the experimentally obtained $T_C$ is much higher than the estimated value.

## 5. Inter-band scattering in $Mo_{1-x}Re_x$ alloys

The existence of phonon assisted s–d scattering can be further confirmed by analysing the temperature dependence of resistivity [29, 49] using the generalized form [50]

$$\rho = \rho_0 + \rho_{ee}T^2 + \rho_{sd}(T/\theta_D)^3 J_3(\theta_D/T) + \rho_{ss}(T/\theta_D)^5 J_5(\theta_D/T), \quad [8]$$

where $J_N(\theta_D/T)$ is given by

$$J_N(\theta_D/T) = \int_0^{\theta_D/T} dz \frac{z^N \exp(z)}{(\exp(z) - 1)^2}. \quad [9]$$

The first term $\rho_0$ is the residual resistivity, $\rho_{ee}$, $\rho_{sd}$ and $\rho_{ss}$ are the coefficients of the temperature dependent resistivity arising from electron–electron, phonon assisted inter-band s–d and intra-band s–s interactions respectively. However, in the range $0.10 \le x \le 0.40$, the $\rho(T)$ shows tendency towards saturation at high temperatures. This behaviour is attributed to the significant contribution from intra-band d–d scattering to resistivity ($\rho_{sat}$) which is temperature independent, and the $\rho(T)$ in the entire temperature range can be accounted for with the help of the following equation:

$$\rho(T)^{-1} = \rho^{-1} + \rho_{sat}^{-1} \quad [10]$$

The temperature dependence of resistivity in several transition metal elements, alloys and compounds has been explained by this generalized form and the contributions to the temperature dependence of resistivity from different mechanisms were estimated by several authors [29, 49, 50]. Figure 9 shows the temperature dependence of normalized resistivity $\rho(T)/\rho(300\ K)$ of the $Mo_{1-x}Re_x$ alloys, revealing the evolution of a subtle hump like structure at about 150 K in $\rho(T)$ when Re is alloyed with Mo. The solid line in figure 9, shows the fit using equations (8), (9) and (10). A $T^5$ variation is predominant in Mo which is related to phonon assisted s–s scattering. As Re is alloyed with Mo, the $T^5$ variation gradually changes to a $T^3$ variation which is related to phonon assisted inter-band s–d scattering. We have observed that the value of $\rho_0$ monotonically increases as a function of x. The electron–electron interaction is observed to be negligible in this system when compared with the other phonon mediated interactions. This fact is also supported by the relatively small values of $\gamma$ in the present system. In figure 10 we compare the various coefficients $\rho_0$, $\rho_{sd}$ and $\rho_{ss}$ used to fit the $\rho(T)$ using equations (8), (9) and (10). It is observed that the intra-band s–s scattering is appreciable only in the samples with $x < x_{C2}$ and is negligible or absent in the samples with $x > x_{C2}$. The inter-band s–d scattering increases with increasing x and becomes appreciable only above $x > x_{C2}$. The value of $\theta_D$ obtained from the fitting of $\rho(T)$ is similar to that obtained from the heat capacity measurements.

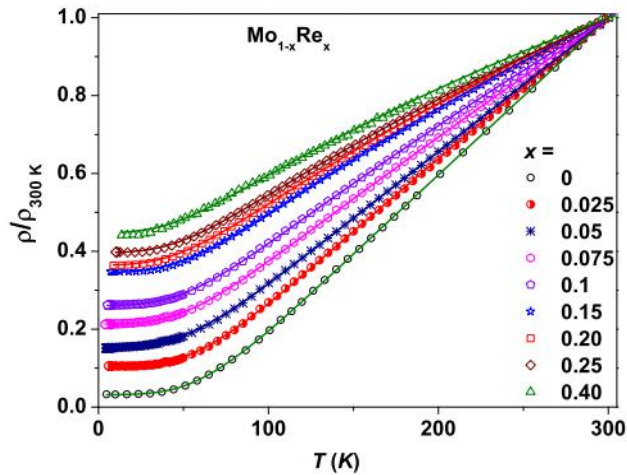

Figure 9. Temperature dependence of normalized resistivity $\rho(T)/\rho(300\ K)$ of the $Mo_{1-x}Re_x$ alloys which shows the evolution of a subtle hump like structure at about 150 K in $\rho(T)$ when Re is alloyed with Mo. The solid lines show the fit using equation (8), (9) and (10). A $T^5$ variation related to s–s scattering in Mo changes to a $T^3$ variation related to phonon assisted s–d scattering in the alloy compositions.

As pointed out earlier, the $Mo_{1-x}Re_x$ system undergoes two ETTs at x = 0.05 and 0.11 [9–15]. The changes in the physical properties is quite drastic across $x_{C2}$ = 0.11 [9]. Our resonant photo-emission spectroscopy and band structure studies [51] reveal that the contribution of 4d like states to the DOS at the Fermi level is quite small in Mo. For alloys with $x > x_{C2}$, the DOS at the Fermi level is dominated by the Re 5d like states. These states are the consequence of the ETT at $x_{C2}$ = 0.11. The angle resolved photo-emission spectroscopy studies by Okada et al show that the newly appeared states are localized along the H–N direction of the Brillouin zone [15]. Thus, we can expect the phonon assisted inter-band s–d scattering to take place between these newly appeared Re 5d like states along the H–N direction of the Brillouin zone and the rest of the Fermi surface which is more like s–p states. Our recent study on the temperature dependence of heat capacity in the superconducting state revealed that the multiband effects are important in x = 0.25 and x = 0.40 [21]. Our conjecture was that both Re 5d like states and s–p like states contribute to the superconductivity in a distinct manner. We have also observed that the contribution of the smaller gap to the superconductivity in x = 0.40 is quite small (about 2%) [21]. We attributed this to the inter-band scattering which has now been confirmed from the present studies. Generally, inter-band scattering driven by impurities results in the lowering of $T_C$ [52]. But the present studies reveal that the inter-band scattering is assisted by phonons. Therefore, electron–phonon coupling constant corresponding to s–d inter-band scattering is enhanced. Recent theoretical works have shown that the phonon mediated inter-band coupling can increase the $T_C$ of multiband superconductors [53–58]. In fact, it has been predicted that superconductivity can occur due to strong inter-band coupling even if the bands as such are not superconducting [53, 54]. Thus, the $T_C$ of the Mo$_{1-x}$Re$_x$ alloys is enhanced in spite of increased inter-band scattering.

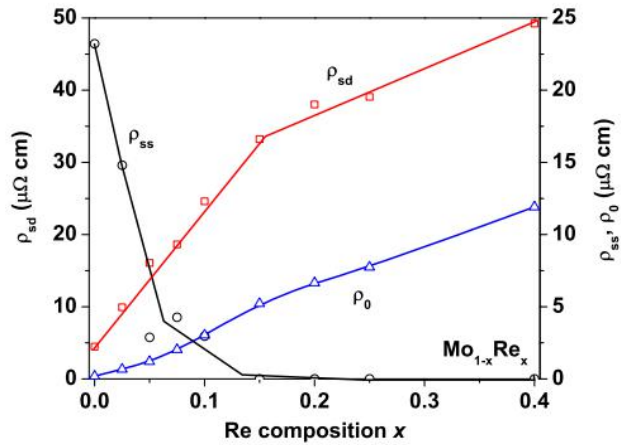

Figure 10. Comparison of coefficients of the resistivity arising from impurities $\rho_0$, phonon assisted inter-band s–d interaction $\rho_{sd}$ and that due to intra-band s–s interaction $\rho_{ss}$ with respect to alloy composition x. The solid lines are guide to eye. The $\rho_0$ monotonically increases as a function of x. For alloys with $x > 0.15$, the $\rho_{ss}$ is zero, where as $\rho_{sd}$ is strongly enhanced as compared to that of Mo.

## Conclusion

In conclusion, we have investigated on the possible reasons for the enhancement of the superconducting transition temperature in the Mo$_{1-x}$Re$_x$ alloys (in comparison with the end elements Mo and Re) through an experimental study of the temperature dependence of normal state resistivity and heat capacity and a theoretical estimation of the electronic DOS obtained with the help of band structure calculation. The superconducting transition temperature $T_C$ and the coefficient of the electronic contributions to the heat capacity $\gamma$ show an abrupt change above $x > 0.1$. The compositional dependence of the electronic DOS at $E_F$ does not follow the compositional dependence of $T_C$ and $\gamma$, thus ruling out the dominant role of DOS in enhancing the $T_C$ of the Mo$_{1-x}$Re$_x$ alloys. The analysis of $C(T)$ supports the lattice softening picture in the Mo$_{1-x}$Re$_x$ alloys. However, the enhancement of $T_C$ above $x = 0.1$ could not be accounted for by this lattice softening alone. The resistivity $\rho$ $(T)$ predominantly varies as $T^5$ in Mo, and changes gradually to a $T^3$ variation in these alloys with increasing Re concentration indicating a changeover from an intra-band

s–s scattering in Mo to a phonon assisted inter-band s–d scattering in the $Mo_{1-x}Re_x$ alloys. We have earlier reported the existence of multi-band effect in the superconductivity of $Mo_{1-x}Re_x$ alloys [21]. Our present analysis suggests that such phonon assisted inter-band s–d scattering in the presence of the multiple bands at the Fermi level can possibly explain the enhancement of $T_C$ in $Mo_{1-x}Re_x$ alloys above $x = 0.1$.

## Acknowledgments

We would like to thank R K Meena for the sample preparation, V S Tiwari, and G Singh for the x ray diffraction measurements and L S Anusha for the help in utilizing the Levenberg–Marquardt algorithm for numerical fitting of the C(T).